\input{aipcheck}

\documentclass[
    ,final            
  ]
  {aipproc}

\layoutstyle{6x9}

\newcommand\lsim{\mathrel{\rlap{\lower4pt\hbox{\hskip1pt$\sim$}}
    \raise1pt\hbox{$<$}}}
\newcommand\gsim{\mathrel{\rlap{\lower4pt\hbox{\hskip1pt$\sim$}}
    \raise1pt\hbox{$>$}}}
\newcommand{\ba}{\begin{array}}
\newcommand{\ea}{\end{array}}
\newcommand{\nn}{\nonumber}

\newcommand{\be}{\begin{equation}}
\newcommand{\ee}{\end{equation}}
\newcommand{\bear}{\begin{eqnarray}}
\newcommand{\eear}{\end{eqnarray}}

\newcommand{\ket}{\,\rangle}
\newcommand{\bra}{\langle \,}

\newcommand{\cO}{{\cal O}}

\newcommand{\mL}{\mathcal{L}}

\newcommand{\mM}{\mathcal{M}}

\newcommand{\Frac}[2]{\frac{\displaystyle #1}{\displaystyle #2}}
\newcommand{\Int}{\displaystyle{\int}}

\begin{document}

\title{Can a resonance theory
\\
be a renormalizable theory?}

\classification{
11.15.Pg,
12.39.Fe,
11.10.Gh
}
\keywords      {$1/N_C$ Expansion, Chiral Lagrangians, Renormalization}

\author{J.J. Sanz-Cillero}{
  address={Department of Physics, Peking
University, Beijing 100871, P.R. China  }
}


\begin{abstract}
In this talk we make an exhaustive analysis of
the possible chiral invariant operators
that may described the resonance decay $S\to\pi\pi$.
These provide at the same time
the only available chiral invariant structures for the loop ultraviolet
divergences in this amplitude. Independently of the order
in perturbation theory, we find just one single-trace term
(four if multi-trace operators are allowed), whose renormalization
renders the matrix element finite.
\end{abstract}

\maketitle


\section{Introduction}

Although there is no argument that allows to affirm that the whole
hadronic action is renormalizable, it is possible to prove this
for some sectors of the theory~\cite{simplify}~\footnote{Talk given at
the International Workshop on QCD, QCD @ Work 2007,
Martina Franca, Italy (16-20 June 2007).
Work done  in collaboration
with L.Y. Xiao.  }.
The motivation of the present work can be found in the
analysis by Rosell {\it et al.} where
the resonance chiral theory
one-loop generating functional was calculated~\cite{generating}.
Only chiral Goldstones, scalars and pseudo-scalar resonances
were considered in the lagrangian of their approach.
All the one-loop ultraviolet (UV) divergences of the theory were computed,
finding the corresponding chiral operators required to fulfill the renormalization.
However, some new operators that could have been {\it a priori}  expected
were  not necessary to render the functional finite. In particular,
a later work~\cite{vanishing} found that,
after imposing a vanishing behaviour at  high-energies,
there were no new UV divergent structures in the
one-loop SS-PP correlator~\footnote{
Defined as
$\Pi(q^2)_{_{SS-PP}}=i \Int dx^d e^{iqx}\bra 0|\mbox{T}\{
J(x)^\dagger J(0) - J_5(x)^\dagger J_5(0)\}|0\ket$,
with $J=\bar{q} q'$, $J_5=i\bar{q}\gamma_5 q'$.
}.
All one needed to make the amplitude finite was a renormalization of
the couplings already in the original lagrangian.
A similar result was found in a dispersive analysis of two-point
Green-functions~\cite{disp-reno}.

These results have provided  some
clues  that may help to understand the way how phenomenological lagrangians
must be  constructed. Thanks to meson field redefinitions
in the generating functional $W[J]$ it is possible
to greatly simplify the structure of
the hadronic action,
with the simplifications occurring  at the level of the lagrangian,
not of particular amplitudes~\cite{generating,georgi,natxo-tesis}.
Once the operators are removed from the action they are no longer relevant
for either on-shell, off-shell, tree-level or loop amplitudes.
This is particularly relevant when the calculation is taken
to the loop level~\cite{generating,natxo-tesis}--\cite{Meissner}.

These techniques are applied to the analysis of the
scalar meson decay into two Goldstones,
$S\to\pi\pi$, which is found to be described at tree-level
by a  finite basis  of  chiral invariant operators.
Several important conclusions are extracted, as
the  fully model-independent description of the $S\pi\pi$--vertex
and the existence of a finite number of local chiral-invariant
structures for the UV loop divergences for this amplitude.
The chiral limit is assumed all along the work.

\section{A chiral theory for resonances}

\subsection{Building blocks of the hadronic action}

We denote as resonance chiral theory (R$\chi$T)
to the most general chiral invariant theory including the Goldstones from
the spontaneous chiral symmetry breaking and the mesonic resonances.
The recovery of chiral perturbation theory
($\chi$PT)~\cite{chpt-Weinberg}--\cite{ChPT-op6}
at low energies requires the R$\chi$T lagrangian to be invariant
under chiral transformations.
This reduces the number of structures that we can build; in general,
just putting  meson fields together
(e.g. $i \lambda \bra  A_{\mu\nu} [ V^{\mu\nu},\pi] \ket$)
does not produce chiral invariant terms
(even if they are invariant under  $SU(3)_V$)
as they may lack of the minimal derivative
structure required in the Goldstone interaction.

The building blocks of the theory are the covariant tensors
made out of the Goldstone fields and $q\bar{q}$ resonance multiplets
$(V,A,S...)$.
The Goldones enter through a non-linear realization
that transforms like
$
(\xi_L(\pi),\xi_R(\pi))
\stackrel{g}{\longrightarrow} \left( g_L \xi_L(\pi) h(g,\pi)^\dagger,
g_R \xi_R(\pi) h(g,\pi)^\dagger \right)
$.
We choose the canonical coset representatives
$\xi_R(\pi)=\xi_L^\dagger(\pi)\equiv u(\pi)$,
with the exponential realization
$u=\exp{\{ i \pi/\sqrt{2}\, F\} }$~\cite{ChPT-op6}.
Combined  together with the external auxiliary fields
$J=s,p,\ell^\mu,r^\mu$ it is the possible to define  the basic tensors
\begin{eqnarray}
u_\mu &=&  i \, \{ u^\dagger (\partial_\mu - i r_\mu) u  \,
-\, u \, (\partial_\mu- i \ell_\mu) u^\dagger \}\, ,
\nn
\\
\chi_\pm &=& u^\dagger \, \chi\, u^\dagger \, \pm \, u\, \chi^\dagger \, u \, ,
\label{eq.bricks}
\\
 f_\pm^{\mu\nu} &=& u\, F_L^{\mu\nu}\, u^\dagger \, \pm \, u^\dagger \, F_R^{\mu\nu}\, u\, ,
 \nn
\end{eqnarray}
which transform covariantly in the form
\begin{equation}
\label{eq.X}
X\,\,\, \stackrel{g}{\longrightarrow}\,\,\,
h\, X\, h^\dagger\, .
\end{equation}
The field $\chi= 2B_0(s+i p)$ contains the scalar and pseudo-scalar
external sources, $s$ and  $p$ respectively.
The $F_{R,L}^{\mu\nu}$  are the strength-field tensors of
the  $r^\mu$ and $\ell^\mu$ sources~\cite{chpt,ChPT-op6}.

The other ingredients of the theory are the
$\bar{q}q$ resonances, which
transform linearly as $U(3)_V$ multiplets under the vector subgroup.
The variation under a general element of the chiral group
is defined by
$R\stackrel{g}{\longrightarrow} h\, R\, h^\dagger$,
similar to that  in Eq.~(\ref{eq.X})~\cite{rcht}.

In addition to the covariant tensors $X=u^\mu,\chi_\pm f_\pm^{\mu\nu},R$,
one can construct terms of the form
$\nabla^\alpha ... \nabla^\mu X$,  with as many covariant derivatives as desired
and also transforming covariantly like in Eq.~(\ref{eq.X}).
The covariant derivative is given by~\cite{ChPT-op6,rcht}
\begin{equation}
\nabla_\mu\, X\, =\, \partial_\mu \, X\, \, +\, \, [\Gamma_\mu\, , \, X]\, ,
\end{equation}
with the chiral connection
$\Gamma_\mu=\frac{1}{2}\{ u^\dagger\, (\partial_\mu - i r_\mu) u
\, +\, u\, (\partial_\mu- i \ell_\mu) u^\dagger \}$.
The commutation
$[ \nabla_\mu ,\nabla_\nu ]  X = [ \Gamma_{\mu\nu} , X ]$
is provided by the tensor
$ \Gamma_{\mu\nu}=
\frac{1}{4} [u_\mu,u_\nu]  -  \frac{i}{2} f_{+\, \mu\nu}$.

Putting these elements together and taking flavour traces
one gets the different chiral-invariant operators
for the R$\chi$T lagrangian~\cite{rcht}, e.g,
$\bra \nabla_\alpha X_1 \, \nabla^\alpha \nabla_\mu X_2 ...\ket $,
${  \bra X_1\ket  \cdot \bra \nabla_\mu X_2 ...\ket    }$...
However, symmetry does not impose any constraint on the number of derivatives
or resonance fields, it only determines the
way how the hadronic fields must be combined ~\cite{rcht}--\cite{donoghue}.
$\bra ...\ket$ is short for trace in the flavour space.

\vspace*{-0.5cm}
\subsection{Challenges in the construction of hadronic lagrangians}

At the moment of writing down a hadronic description of QCD
there is a set of important issues that must be addressed.
First, one needs a formal perturbation theory on some
parameter that suppresses hadron loops and makes lowest order contributions dominant.
The $1/N_C$ expansion based on
't Hooft's large number of colours limit~\cite{NC1}
seems to be the most suitable one for a theory with resonances~\cite{natxo-tesis},
being each meson loop suppressed by a power of $1/N_C$~\cite{NC3}.

The validity of the  $1/N_C$ expansion for any energy allows to connect the
resonance theory with the high-energy description of QCD, provided by
perturbative QCD and the operator product expansion (OPE)~\cite{OPE}.
The short-distance matching has produced  very successful  determinations both at
leading order (LO)~\cite{PI:02} and at next-to-leading order
in $1/N_C$~\cite{natxo-tesis,SSPPrcht}, although the uncertainties in the
matching procedure are not yet fully understood~\cite{PADE}.
In any case, no high-energy  analysis is considered in the present work~\cite{simplify},
though it may result fruitful in future studies.

In this talk  we rather
focus on the implementation of chiral symmetry  on the hadronic action.
This allows the construction of an infinite number of invariant operators,
what may look discouraging. However,
one must keep in mind  the real goal:
\begin{itemize}

\item{}
{\bf The action may contain an infinite number of operators.} This is something
already familiar to us through $\chi$PT where even at LO we have an infinite number
of (related) terms due to the non-linear Goldstone realization.

\item{}
Nonetheless, {\bf the crucial point to make the theory predictive is that
for a given amplitude at a given order in the established perturbative expansion,
only a finite number of operators is required.}
This is what happens in the previous example of $\chi$PT. Even if there is an infinity
of terms, only two operators are required to describe, for instance, the
$\pi\pi$ scattering at LO in the chiral expansion.

\end{itemize}

\section{Simplifying the R$\chi$T lagrangian}

\subsection{Intuitive picture, formal procedure}

\begin{figure}
\includegraphics[height=3.3cm,clip]{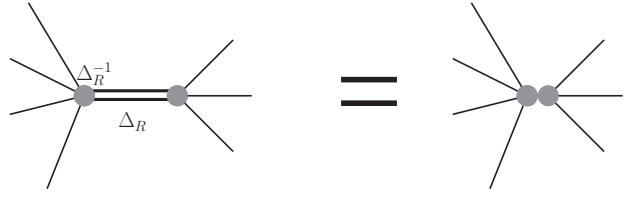}
\caption{
Local structure in non-local meson exchanges.}
\label{fig.propa-contract}
\end{figure}

Some operators of the R$\chi$T lagrangian, allowed by the symmetry, are actually
redundant and do not carry any extra physical information.
An intuitive way to understand this
relies on the picture in Fig.~(\ref{fig.propa-contract}).
The contribution from some operators
may look like a non-local meson exchange but, nevertheless, they always appear
through local structures. Due to the form of the vertex (for instance,
$\lambda \bra ... (\nabla^2 +M_S^2) S\ket $),
the intermediate propagator is canceled out any time this vertex enters into play,
resembling a local operator contribution.
Since R$\chi$T includes all the operators compatible with the symmetry,
the structure on the right-hand side of Fig.~(\ref{fig.propa-contract})
is already contained in the lagrangian.

The formal procedure to remove these redundant terms relies on the freedom
to perform meson field redefinitions in the generating functional $W[J]$~\cite{georgi}.
This transforms some operators into others and, if it
is conveniently tuned,  it is possible to fully remove
the undesired terms.

\vspace*{-0.5cm}
\subsection{Goldstone and scalar resonance transformations}

The starting point is the R$\chi$T lagrangian, with
the completely general structure~\cite{simplify}
\begin{eqnarray}
\mL &=& \Frac{F_0^2}{4}\bra u_\mu u^\mu\ket \,
+\, \bra A_S\, \nabla^\mu u_\mu\ket \,
+\, \bra B_S\ket
\, -\, \Frac{1}{2}\bra S (\nabla^2 +M_S^2) S\ket
\, +\, \Delta \mL
\, ,
\label{eq.La}
\end{eqnarray}
where $\Delta\mL$ is not relevant in the present study
and we just provide its general structure:
\begin{eqnarray}
\Delta\mL &=&  \cO(S^2 u^\alpha u^\beta) \, + \cO(S^{3})
\,  +\, \cO( R')
\, + \, \cO(J)\, +\,\cO(u^\alpha u^\beta u^\mu u^\nu)\,  ,
\label{eq.DLa}
\end{eqnarray}
with the term $\cO(R')$ containing at least one resonance
$R'\neq S$.  $\bra A_S \nabla^\mu u_\mu\ket$ and $\bra B_S\ket$
account for all the operators made out of just one $S$--meson field
and two tensors $u^\alpha$, but allowing
any number of  covariant derivatives:
$A_S \nabla^\mu u_\mu, \, B_S\,
\sim\,  S \, \nabla ...u^\alpha\,  \nabla ...u^\beta $.

We will perform first a Goldstone field redefinition that induces
a shift in $u^\mu$ of the form
$ u_\mu \longrightarrow  u_\mu  +\frac{2}{F_0^2}\, \nabla_\mu A_S
+ \cO(A_S^2)$~\cite{simplify}.
The required Goldstone transformation is not unique, being one of the simplest
${   \xi_R\to \xi_R \exp{\{ - i A_S/F_0^2\} }  ,    }  \,\,
   {   \xi_L\to \xi_L \exp{\{ i A_S/F_0^2 \} }     }$.
This produces a lagrangian with exactly
the same structure as in Eqs.~(\ref{eq.La})--(\ref{eq.DLa})
except for  the term $\bra A_S\nabla^\mu u_\mu\ket$, which is completely removed.

The second step relies on the scalar resonance field transformation.
The remnant term $\bra B_S\ket$ is decomposed in the form,
\begin{eqnarray}
\bra B_S\ket \,\, &=&\,\,\bra \zeta \, (\nabla^2 +M_S^2) S \ket
\, +\, \bra \eta \,  S  \ket \, ,
\label{eq.BS}
\end{eqnarray}
where $\zeta$ and $\eta$ are local chiral tensors containing just Goldstones.
At this point it is easy to realise that the change
$S\to S +\zeta$  removes the first term on the right-hand side of
Eq.~(\ref{eq.BS}), leaving the finally simplified lagrangian,
\begin{eqnarray}
\mL &=&
\Frac{F_0^2}{4}\bra u^\mu \, u_\mu\ket  \, +\, \bra \eta \, S \ket \,
- \, \Frac{1}{2}\bra S (\nabla^2+M_S^2) S\ket \, +\, \Delta \mL
\, ,
\label{eq.Lb}
\end{eqnarray}
where all the operator that could be written like
$\bra A_S\nabla^\mu u_\mu\ket $ or $\bra \zeta\,  (\nabla^2 +M_S^2) S \ket$
have been removed.

\vspace*{-0.5cm}
\subsection{$S\to\pi\pi$ decay amplitude}

In the construction of
the most general form for chiral invariant operators contributing to $S\to\pi\pi$
we have to take into account that, in the chiral limit, we cannot
include the tensors $\chi_\pm$, $f_\pm^{\mu\nu}$ since they are proportional
to external sources.   We must include one $S$ field
and exactly two tensors $u^\alpha$. Otherwise, the operator does not preserve parity
or it produces more than two Goldstones in the final state.
No {\it a priori}   restriction can be made on the number of covariant derivatives.
In the chiral limit, this gives the general form
\begin{eqnarray}
\mL_{S\to\pi\pi}
\, &=& \, \lambda \, \bra S\, \left\{\, \nabla^{\mu_1}\, ...\, \nabla^{\mu_m}\, u^\rho\, ,
\, \nabla^{\nu_1}\, ...\, \nabla^{\nu_n}\, u^\sigma \right\}\ket
 \,\,\times
\, \, t_{\mu_1, ...\, \mu_m, \, \rho, \, \nu_1,...\, \nu_n,\, \sigma}\quad ,
 \label{eq.general-Spipi}
\end{eqnarray}
where the Lorentz tensor
$t_{\mu_1, ...\, \mu_m, \, \rho, \, \nu_1,...\, \nu_n,\, \sigma}  $
handles all the possible contractions of the indices. The anticommutator $\{...\, ,\, ...\}$
ensures that the operator is invariant under charge and hermitian
conjugations~\cite{ChPT-op6}.

The simplest operator of this kind is the familiar term,
\begin{eqnarray}
\mL_{S\to \pi\pi}\, \, =\, \, \lambda \, \bra \, S\, \{ u^\mu \, , u_\mu \}\, \ket
\,\, =\,\,  2\, \lambda \bra S\,  u^\mu \, u_\mu\ket \, ,
\label{eq.cd}
\end{eqnarray}
which is just the $c_d \bra S u^\mu u_\mu \ket$ operator in Ref.~\cite{rcht}.

For a higher number of derivatives,
one has different possible contractions of the Lorentz indices.
The detailed analysis of the different cases is done
in Ref.~\cite{simplify}, where it is concluded that these operators
either show the structure $\bra A_S\nabla^\mu u_\mu\ket$, or the form
$\bra \zeta \, (\nabla^2 +M_S^2)S\ket$, or  they are equivalent to
an operator with two derivatives less. The first two correspond to operators that can
be fully removed through field redefinitions and the third one allows
to iteratively simplify the operator and to reduce it into the
$c_d$ term in Eq.~(\ref{eq.cd}).

If multitrace  operators -subleading in $1/N_C$- are allowed then there are another
three independent operators:
$\lambda_a \bra S\ket \, \bra u_\mu u^\mu\ket $,
$\lambda_b \bra S u_\mu \ket \, \bra u^\mu\ket $
and $\lambda_c \bra S\ket \, \bra u_\mu\ket \, \bra u^\mu\ket $,
exhausting the list of chiral invariant operators contributing to the
decay $S\to \pi\pi$.

\section{Finite basis and renormalizability}

This provides a clear example of the possibility of constructing a fully model independent
lagrangian for the description of hadronic processes.
As it has been noted, the action may contain an infinite number of operators
but the $S\to\pi\pi$ amplitude is given at large--$N_C$ by just the $c_d$ term.

What implications does this have on the renormalizability of the
amplitude? The only available local chiral invariant structures
for the UV divergences appearing in the $S\to\pi\pi$ decay at the loop level
are these four operators
$O_{c_d}, O_{\lambda_a},O_{\lambda_b},O_{\lambda_c}$.
Therefore, the renormalization of
the four couplings $c_d$, $\lambda_a$, $\lambda_b$, $\lambda_c$ renders
the amplitude finite at any order in perturbation theory.

Preliminary studies have found similar simplifications in a wider set of
amplitudes, which are also described at tree-level by a finite number of
independent operators~\cite{inprogress}.
We plan to extend the analysis to  other $S$--meson processes and amplitudes with other
resonances. It can be also applied to the heavy quark meson sector and to the study
of Green-functions.

The possibility of this to be a general feature of the lagrangian is
more difficult to defend although it would lead to very deep implications.
If an amplitude $\mM$ at any order in perturbation theory is
given at tree-level by a fixed and finite number $N$ of chiral invariant operators
($S[\phi]=...+ \sum_{k=1}^N  \Int dx^D   c_k O_k[\phi,J]$)
then the local UV divergences can only show this structure
and the generating functional has then the form
\begin{equation}
W[J] \, = \, ... \,  + \,  \sum_{k=1}^N\Int dx^D  c_k  O_k[\phi^{cl},J]
\, \, \,
+\,\,\,
\sum_{k=1}^N\Int dx^D  \lambda_\infty \, \Gamma_k  O_k[\phi^{cl},J]
\, \, \, ,
\end{equation}
with $\lambda_\infty$ containing the UV divergence. Thus, the renormalization
of the couplings $c_k$ would render the amplitude finite at any order
in perturbation theory.

\begin{theacknowledgments}
I would like to thank the organisers of QCD@Work 2007 for their kindness and
the attentions received during the conference.
 This work is supported in part by National
 Nature Science Foundations of China under contract number
 10575002,
 10421503.

\end{theacknowledgments}

\bibliographystyle{aipproc}   

\end{document}